\documentstyle[twoside,fleqn,espcrc2,psfig]{article}

\newcommand{\PL}{{\em Phys.\ Lett.\ }}

\newcommand{\tr}{{\rm Tr}}

\newcommand{\norm}[1]{\raise.3ex\hbox{:}#1\raise.3ex\hbox{:}}

\title{D-branes and Creation of Strings
}

\author{Igor R. Klebanov%
\address{ Joseph Henry Laboratories \\ 
          Princeton University \\ 
          Princeton, New Jersey 08544}
}

\begin{document}

\begin{abstract}
We review two types of D-branes processes where open strings are
created. In the first type, a closed string incident on a collection
of D-branes is converted into a number of open strings running along
them. For the case of threebranes we compare the leading absorption rate
with that in semiclassical gravity, and find exact agreement.
A supersymmetric non-renormalization theorem guarantees that this
agreement survives all corrections in powers of the string coupling
times the number of branes.
The second type of process is creation of stretched open strings by
crossing D-branes. We show that this is possible whenever a $p$-brane
passes through an $(8-p)$-brane positioned orthogonally to it.
The extra attractive force exerted by the stretched open string
is crucial for finding that the net force cancels in this BPS system.

\end{abstract}

\maketitle


\newcommand{\beq}{\begin{equation}}
\newcommand{\eeq}{\end{equation}}
\newcommand{\beqs}{\begin{eqnarray}}
\newcommand{\eeqs}{\end{eqnarray}}
\newcommand{\laa}{\lambda_{IIA}}
\newcommand{\lop}{\lambda_{I'}}
\newcommand{\lhet}{\lambda_{E_8}}
\newcommand{\rop}{R_{I'}}
\newcommand{\rhet}{R_{E_8}}
\newcommand{\da}{{\dot a}}
\newcommand{\db}{{\dot b}}




\renewcommand{\epsilon}{\varepsilon}
\def\fixit#1{}
\def\comment#1{}
\def\equno#1{(\ref{#1})}
\def\equnos#1{(#1)}
\def\sectno#1{section~\ref{#1}}
\def\figno#1{Fig.~(\ref{#1})}
\def\D#1#2{{\partial #1 \over \partial #2}}
\def\df#1#2{{\displaystyle{#1 \over #2}}}
\def\tf#1#2{{\textstyle{#1 \over #2}}}
\def\d{{\rm d}}
\def\e{{\rm e}}
\def\i{{\rm i}}
\def\Leff{L_{\rm eff}}

\def\vp{{\bf p}}
\def\al{\alpha}
\def\ab{\bar{\alpha}}
\def \bi{\bibitem}
\def \ep{\epsilon}
\def\D{\Delta}
\def \om {\omega}
\def\LL{\td \l}
\def \do {\dot}
\def\H {{\cal H}}
\def \B {{\cal B}}
\def \ua {\uparrow}
\def \Q {{\hat Q}}
\def \P {{\hat P}}
\def \q {{\hat q}}
\def \bp{{\bar \psi}}

\def \k {\kappa} 
\def \F {{\cal F}}
\def \g {\gamma}
\def \del {\partial}
\def \bd {\bar \partial }
\def \na {\nabla}
\def \const {{\rm const}}
\def \na {\nabla }
\def \D {\Delta}
\def \a {\alpha}
\def \b {\beta}
\def\r {\rho}
\def \s {\sigma}
\def \p {\phi}
\def \m {\mu}
\def \n {\nu}
\def \vp {\varphi }
\def \l {\lambda}
\def \t {\tau}
\def \td {\tilde }
\def \ci {\cite}
\def \sm {$\s$-model }

\def \o {\omega}
\def \inv {^{-1}}
\def \ov {\over }
\def \four{{\textstyle{1\over 4}}}
\def \fourth{{{1\over 4}}}
\def \ha {{1\ov 2}}
\def \QQ {{\cal Q}}


\def\comment#1{}
\def\fixit#1{}

\def\tf#1#2{{\textstyle{#1 \over #2}}}
\def\df#1#2{{\displaystyle{#1 \over #2}}}

\def\coth{\mathop{\rm coth}\nolimits}
\def\csch{\mathop{\rm csch}\nolimits}
\def\sech{\mathop{\rm sech}\nolimits}
\def\Vol{\mathop{\rm Vol}\nolimits}
\def\vol{\mathop{\rm vol}\nolimits}
\def\diag{\mathop{\rm diag}\nolimits}
\def\tr{\mathop{\rm Tr}\nolimits}

\def\sqr#1#2{{\vcenter{\vbox{\hrule height.#2pt
         \hbox{\vrule width.#2pt height#1pt \kern#1pt
            \vrule width.#2pt}
         \hrule height.#2pt}}}}
\def\square{\mathop{\mathchoice\sqr34\sqr34\sqr{2.1}3\sqr{1.5}3}\nolimits}

\def\TL{\hfil$\displaystyle{##}$}
\def\TR{$\displaystyle{{}##}$\hfil}
\def\TC{\hfil$\displaystyle{##}$\hfil}
\def\TT{\hbox{##}}

\def\shortlistrefs{\footatend\bigskip\bigskip\bigskip%
Immediate\closeout\rfile\writestoppt
\baselineskip=14pt\centerline{{\bf References}}\bigskip{\frenchspacing%
\parindent=20pt\escapechar=` Input refs.tmp\vfill\eject}\nonfrenchspacing}

\def\eff{{\rm eff}}
\def\abs{{\rm abs}}
\def\hc{{\rm h.c.}}
\def\+{^\dagger}

\def\cl{{\rm cl}}

\def\M{\cal M}
\def\D#1#2{{\partial #1 \over \partial #2}}

\def\overleftrightarrow#1{\vbox{Ialign{##\crcr
     \leftrightarrow\crcr\noalign{\kern-0pt\nointerlineskip}
     $\hfil\displaystyle{#1}\hfil$\crcr}}}

\def \t {\tau}
\def \td {\tilde }
\def \ci {\cite}
\def \sm {$\s$-model }

\def \o {\omega}
\def \inv {^{-1}}
\def \ov {\over }
\def \four{{\textstyle{1\over 4}}}
\def \fourth{{{1\over 4}}}
\def \ha {{1\ov 2}}
\def \QQ {{\cal Q}}

\def \lr { \lref}
\def\np {{  Nucl. Phys. }}
\def \pl {{  Phys. Lett. }}
\def \mpl {{ Mod. Phys. Lett. }}
\def \prl {{  Phys. Rev. Lett. }}
\def \pr  {{ Phys. Rev. }}
\def \ap  {{ Ann. Phys. }}
\def \cmp {{ Commun.Math.Phys. }}
\def \ijmp {{ Int. J. Mod. Phys. }}
\def \jmp {{ J. Math. Phys.}}
\def \cqg {{ Class. Quant. Grav. }}


\section{Introduction}

The Dirichlet branes (D-branes) \cite{dlp,polch}
are a remarkable window into non-perturbative
string theory. Their dynamics is neatly described
by open strings whose end-points are free to
move only along a $p$-dimensional
hyperplane, the center of the Dirichlet $p$-brane.
Since they are exchanged with the elementary strings under
non-perturbative duality symmetries, the D-brane degrees of freedom
are indispensable for the overall consistency of the theory.
There are nice reviews explaining the physics of D-branes and
their applications to string dualities \cite{reviews}.

While D-branes are non-perturbative objects whose tensions scale
as $1/g_{\rm str}$, their interactions with closed strings 
are tractable in perturbation theory \cite{akirev}.
An example of such an interaction is scattering of a massless closed
string off a D-brane, which may be used to measure the D-brane
form factors
\cite{KThor}. The leading order amplitudes of this type involve two
bulk vertex operators on a disk and were calculated in 
\cite{KThor,GHKM,gm,akirev}. 

Another example is absorption of
a closed string accompanied by production of some number of
open strings moving along the D-brane. For general incoming energy,
the lowest order in perturbation theory involves two open strings
in the final state. The necessary amplitudes, which involve one bulk
and two boundary vertex operators, were calculated explicitly in
\cite{HK,akirev}. While at high incident energy they were found to
fall-off exponentially, which indicates growth in the effective
thickness of a D-brane, at low energies these amplitudes are in full
agreement with the effective actions of the 
Dirac-Born-Infeld type \cite{dlp,reviews}.
Thus, the low-energy behavior of the absorption cross-sections for
massless particles may be reliably determined from the world
volume effective actions of the D-branes. It is of obvious interest to
compare the results with the absorption cross-sections by the
$p$-brane solutions in low-energy supergravity. This was
done in \cite{IK,GKT} and will be reviewed in section 2. For the
threebrane, which is the only non-singular D-brane solution,
a perfect agreement of the leading cross-sections was 
found \cite{IK,GKT}. The higher order processes where $\ell+2$
open strings are produced by an incoming closed string turn out
to be counterparts of the semiclassical
absorption in the $\ell$-th partial wave. Thus,
interesting comparisons can also be made for the higher partial waves.
We also compare the semiclassical and
the world volume absorption cross-sections for the twobrane
and the fivebrane of M-theory, and find agreement in the scaling with
the energy. The comparison of the coefficients, however, is impeded
by the lack of information about the world volume theory
of multiple M-branes.

In section 3 we move on to another aspect of the D-brane physics.
We study the force balance between orthogonally
positioned $p$-brane and $(8-p)$-brane.
The force due to graviton and dilaton exchange is repulsive in this
case. We identify the attractive force that balances this repulsion
as due to one-half of a fundamental string stretched between the
branes \cite{DFK}. If the $p$-brane is initially to the left of the
$(8-p)$-brane and the half-string points away from the $p$-brane then,
after it crosses to the right, we find that the half-string points toward
the $p$-brane.
This may be interpreted
as creation of one fundamental string directed
from the $(8-p)$-brane to the $p$-brane. We show this directly
from the structure of the Chern-Simons terms in the D-brane effective
actions. We also discuss the effect of string creation on the
0-brane quantum mechanics in the type I' theory.
The creation of a fundamental string is related by
U-duality to the creation of a 3-brane 
discussed by Hanany and Witten \cite{HW}.
Both processes have a common origin in M-theory: as two M5-branes
with one common direction cross, a M2-brane stretched between them is
created.

\section{World Volume Approach to Absorption by Non-dilatonic Branes}

Extremal black holes with non-vanishing horizon area may be embedded
into string theory or M-theory using intersecting $p$-branes
\cite{CTT,sv,cm,mst,myers,at,KT,bl}. 
These configurations are useful for
a microscopic interpretation of the Bekenstein-Hawking entropy.
The dependence of the entropy on the charges and the non-extremality
parameter suggests a connection with $1+1$ dimensional conformal
field theory. This `effective string' is essentially the intersection
of the $p$-branes, but its winding number grows
as the product of the numbers of the branes
involved in the intersection \cite{jl,KT}. Therefore,
the energy of the lowest excitation scales as the inverse of this
product, in agreement with semiclassical considerations \cite{jl}.
Thus, the embedding of black holes into string or M-theory
gives an appealing picture of their lowest-energy excitations.

Calculations of emission and absorption rates 
\cite{cm,dmw,dm,GK,mast,GKtwo,CGKT,KM,hawk,dkt,KK,KRT,Mathur,Gubser,CL} 
provide further tests of the 
`effective string' models of $D=5$ black holes
with three charges and of $D=4$ black holes with four charges.
For minimally coupled scalars the functional dependence of the
greybody factors on the frequency agrees exactly with semiclassical
gravity, providing a highly non-trivial verification of the effective
string idea \cite{mast,GKtwo}. Similar successes have been achieved
for certain non-minimally coupled scalars, which were shown to couple 
to higher dimension operators on the effective string.
For instance, the `fixed' scalars \cite{Kallosh} 
were shown to couple to operators of dimension $(2,2)$ 
\cite{CGKT},
while the `intermediate' 
scalars \cite{KRT} -- to operators of dimension $(2,1)$
and $(1,2)$. Unfortunately, there is little understanding of
the `effective string' from first principles, and some of the
more sensitive tests reveal this deficiency. For instance, 
the semiclassical
gravity calculations of the `fixed' scalar absorption rates
for general black hole charges reveal a gap in our understanding
of higher dimension operators \cite{KK}. 
A similar problem occurs when one attempts
a detailed effective string 
interpretation of the higher partial waves of a
minimally coupled scalar \cite{Mathur,Gubser}. Even the s-wave absorption 
by black holes with general charges is complex enough
that it is not reproduced by the simplest effective string 
model \cite{KM,CL}. These difficulties 
by no means invalidate the general qualitative picture,
but they do pose some interesting challenges. 
We feel that, to gain insight into
the general relation between gravity and Yang-Mills
theory, it is useful to study,
in addition to the intersecting branes,
the simpler configurations
which involve parallel branes only. Apart from their intrinsic interest,
such configurations are also relevant to Schwarzschild black holes in
Matrix theory \cite{BFSS}, as was recently explained in \cite{BFKS,KS}.

A microscopic interpretation of
the entropy of near-extremal $p$-branes was first explored in
\cite{GKP,kt}. It was found that the scaling of the 
Bekenstein-Hawking entropy with the temperature agrees
with that for a massless gas in $p$ dimensions only for the
`non-dilatonic $p$-branes':
namely, the self-dual 3-brane of the type IIB theory,
and the 2- and 5-branes of M-theory.
In \cite{HP} a way of reconciling the differing scalings 
for the dilatonic branes was proposed.
According to this `correspondence principle' \cite{LS,HP}, 
the string theory and the 
semiclassical gravity descriptions are in general
expected to match only at a special value of the temperature, which
corresponds to the horizon curvature 
comparable to the string scale.\footnote{
For $N$ parallel D-branes, $N g_{\rm str}$ is of order 1 at the
matching point \cite{HP}.}
In \cite{HP} it was shown that, in all known cases, the stringy and the
Bekenstein-Hawking entropies match at this point up to factors of order
1. Part of the ambiguity in this factor comes from knowing the
matching point only approximately. However, for the non-dilatonic
branes this ambiguity is absent: the matching can be achieved at any
scale because the stringy and the semiclassical entropies have
identical scalings with temperature. This still leaves 
a relative factor of 4/3  
for the 3-brane entropy \cite{GKP} which, we hope, will eventually
find an exact explanation in terms of the strongly coupled world volume
theory.

The non-dilatonic branes have a number of special properties.
A notable property of their extremal metrics
is that the transverse part of the 
geometry is non-singular: instead of a singularity we find an infinitely
long throat whose radius grows with the charge 
(the vanishing of the horizon area is due to the longitudinal
contraction). 
Thus, for a large number $N$
of coincident branes, the curvature may be made arbitrarily small in
Planck units. For instance, for $N$ D3-branes, the curvature is bounded
by a quantity of order
\beq
{1\over \sqrt{N\kappa_{10}}} \sim {1\over \alpha'\sqrt{N g_{\rm str}}}
\ . 
\eeq
Thus, to suppress the string scale corrections to the classical metric,
we need to take the limit $N g_{\rm str}\rightarrow\infty $.

The tensions of non-dilatonic branes
depend on $g_{\rm str}$ and
$\alpha'$ only through the gravitational constant $\kappa$ in 
the appropriate dimension, which is also the only scale present in the
semiclassical description. Indeed, the D3-brane tension is
\beq
T_{(3)}={\sqrt \pi \over\kappa_{10}}\ , 
\eeq
the M2-brane tension is 
\beq
T_{(2)}=(2 \pi^2 \kappa_{11}^{-2})^{1/3}\ ,
\eeq
and the M5-brane tension is  
\beq
T_{(5)}=\left ({\pi\over 2 \kappa_{11}^4}\right )^{1/3}\ .
\eeq
This suggests that we can compare the expansions of various 
quantities in powers of $\kappa$ between the microscopic and
the semiclassical descriptions. 
For instance, in the 3-brane
absorption cross section the expansion parameter is \cite{IK}
\beq
\label {param}
N\kappa_{10} \omega^4 \sim N g_{\rm str} \alpha'^2 \omega^4
\ ,
\eeq
where $\omega$ is the incident energy. Thus, we may consider
a `double scaling limit'
\beq
\label{ dsl}
 N g_{\rm str}\rightarrow\infty\ ,
\qquad \omega^2 \alpha'\rightarrow 0\ ,
\eeq
where the expansion parameter (\ref{param}) is kept small.
The semiclassical absorption cross section is naturally
expanded in powers of $\omega^4\times {\rm curvature}^{-2}$, which
is the same expansion parameter (\ref{param}) as the one governing the
string theoretic description of the 3-branes. 
The two expansions
of the cross-section thus may indeed be compared, and
the leading term agrees exactly \cite{IK,GKT}. This
provides strong evidence in favor of
absorption by extremal threebranes being a unitary process.
While in the classical calculation the information carried by the
incident scalar 
seems to disappear down the infinite throat of the classical
solution, the stringy approach indicates that the information is not
lost: it is stored in the quantum state of the back-to-back massless
gauge bosons on the world volume which are
produced by the scalar. Subsequent decay of the threebrane
back to the ground state proceeds via annihilation of the gauge
bosons into an outgoing massless state, and there seems to be no
space for information loss.

$N$ parallel D3-branes
are known to be described by a $U(N)$ supersymmetric
gauge theory on the world volume \cite{EW}.
The situation is not as simple in M-theory, where the effective actions
of multiple branes are not known in detail.
In \cite{IK} the scalings of the classical absorption cross-sections
were found to agree with the twobrane and fivebrane effective action
considerations.
The absorption cross-section
of a longitudinally polarized graviton by a single fivebrane
was calculated in \cite{GKT}.
From the world-volume  effective action it was found that the
absorption cross-section is $1/4$ of the rate formally predicted
by classical gravity. In fact, one could hardly expect perfect
agreement for a single fivebrane -- the classical description is
expected
to be valid only for a large number of coincident branes.
It is interesting, nevertheless, how close the two calculations come
to agreeing with each other. A similar comparison
for a single M2-brane reveals a factor
discrepancy of $3\pi/(4\sqrt 2)$ \cite{GKT}.

\subsection{Semiclassical Absorption by Extremal Branes}

In this section we carry out semiclassical absorption calculations for the
three cases of interest: the 3-brane in $D=10$, and the 2- and
5-branes in $D=11$. We study the minimally coupled massless
scalars whose propagation is governed by
\beq
\nabla^\mu \nabla_\mu \phi =0\ .
\eeq

The extremal 3-brane metric \cite{hs} can be written as  
\beqs ds^2 &=& \left (1+{R^4\over r^4}\right )^{-1/2} 
\left (- dt^2 +dx_1^2+ dx_2^2+ dx_3^2\right )
\nonumber \\ &+& \left (1+{R^4\over r^4}\right )^{1/2} 
\left ( dr^2 + r^2 d\Omega_5^2 \right )\ .
\eeqs
The s-wave of a minimally coupled massless scalar satisfies
\beq
\label{Coul} 
\left [\rho^{-5} {d\over d\rho} \rho^5 {d\over d\rho} +
1 + {(\omega R)^4\over \rho^4} \right ] \phi(\rho) =0\ ,
\eeq
where $\rho = \omega r$. Thus, we are interested in absorption by
the Coulomb potential in 6 spatial dimensions. 
For small $\omega R$ this
problem may be solved by matching an approximate
solution in the inner region to an approximate solution
in the outer region.

To approximate in the inner region, it is convenient to
use the variable $z=(\omega R)^2/\rho$. Then 
(\ref{Coul}) turns into
\beq
\label{Coulone} 
\left [{d^2\over d z^2} - {3\over z} {d\over dz}
+1 + {(\omega R)^4\over z^4} \right ] \phi =0\ .
\eeq
Substituting $\phi = z^{3/2} f(z)$, we find
\beq
\label{Coultwo} 
\left [{d^2\over d z^2} - {15\over 4 z^2} 
+1 + {(\omega R)^4\over z^4} \right ] f =0\ .
\eeq
The last term may be ignored if $z\gg (\omega R)^2$, i.e. if
$\rho \ll 1$. In this region, (\ref{Coultwo}) is easily solved in terms of
cylinder functions. Since we are interested in the incoming wave for
small $\rho$, the appropriate solution is
\beq
\label{inner}
\phi = i z^2 
\left [J_2 (z)+
i N_2 (z)\right ] \ , 
\eeq
where $J$ and $N$ are the Bessel and Neumann functions.

Another way to manipulate (\ref{Coul}) is by substituting
$\phi = \rho^{-5/2} \psi$, which gives
\beq
\label{Coulthree}
[{d^2\over d \rho^2} - {15\over 4 \rho^2} 
+1 + {(\omega R)^4\over \rho^4}] \psi =0\ .
\eeq
Now the last term is negligible for $\rho \gg (\omega R)^2$,
where (\ref{Coulthree}) is solvable in terms of cylinder functions.
If $\omega R\ll 1$, then the inner region ($\rho\ll 1$) overlaps
the outer region ($\rho \gg (\omega R)^2$), and the 
approximate solutions may be matched. We find that (\ref{inner})
matches onto
\beq
\label{outer}
\phi = {32\over \pi}
\rho^{-2} J_2(\rho) \ , \qquad \rho \gg (\omega R)^2\ .
\eeq
The absorption probability may be calculated as the
ratio of the flux at the throat to the incoming flux at
infinity, with the result
\beq {\cal P}= {\pi^2\over 16^2} (\omega R)^8 \ .
\eeq
In $d$ spatial dimensions, the absorption cross-section is
related to the s-wave absorption probability by \cite{dgm}
$$ \sigma = {(2 \pi)^{d-1}\over \omega^{d-1} \Omega_{d-1}} {\cal P}
\ ,$$
where
$$\Omega_D = {2 \pi^{{D+1\over 2}}\over \Gamma \left (
{D+1\over 2}\right ) }
$$ 
is the volume of a unit $D$-dimensional sphere.
Thus, for the 3-brane we find\footnote{By absorption cross section
we will consistently mean
the cross section per unit longitudinal volume of the
brane.} \cite{IK}
\beq
\label{three}
\sigma_{\rm 3\ class}= {\pi^4\over 8}\omega^3 R^8 \ . 
\eeq
The  scale  parameter $R$  of the classical 3-brane solution  
is related \cite{GKP,IK} to the number $N$ of coinciding microscopic 
3-branes by the equation 
\beq
R^4 = {\kappa_{10}\over 2\pi^{5/2}} N \ , 
\eeq
which follows from the quantization of the threebrane charge.
Hence, 
\beq \label{classical}
\sigma_{\rm 3\ class}={\kappa_{10}^2  \omega^3 N^2\over 32 \pi}\ .
\eeq

This exercise may be easily repeated for the other two
non-dilatonic branes.
For the M5-brane the extremal metric is \cite{guv}
\beqs ds^2 & = & \left (1+ {R^3\over r^3}\right )^{-1/3} 
\left (-dt^2 +dx_1^2+\ldots +  dx_5^2\right )
\nonumber \\  & + & \left (1+ {R^3\over r^3}\right )^{2/3} 
\left ( dr^2 + r^2 d\Omega_4^2 \right )
\ .\eeqs
Now the s-wave problem reduces to
absorption by the Coulomb potential in 5 spatial dimensions,
$$ \left [\rho^{-4} {d\over d\rho} \rho^4 {d\over d\rho} +
1 + {(\omega R)^3\over \rho^3}\right ] \phi(\rho) =0\ .
$$
A matching calculation gives \cite{IK}
\beq
\label{five}
\sigma_{\rm 5\ class}= {2 \pi^3\over 3}\omega^5 R^9=
{\kappa_{11}^2 N^3 \omega^5\over 2^6 \cdot 3 \pi^2} \ , 
\eeq
where $N$ is the number of coincident M5-branes.
 
For the M2-brane the extremal metric is \cite{ds}
\beqs ds^2 & = &\left (1+ {R^6\over r^6}\right )^{-2/3} 
\left (-dt^2 +dx_1^2+ dx_2^2\right )
\nonumber \\  & + & \left (1+ {R^6\over r^6}\right )^{1/3} 
\left ( dr^2 + r^2 d\Omega_7^2 \right )\ .
\eeqs
Now the s-wave problem reduces to
absorption by the Coulomb potential in 8 spatial dimensions,
$$ \left [\rho^{-7} {d\over d\rho} \rho^7 {d\over d\rho} +
1 + {(\omega R)^6\over \rho^6}\right ] \phi(\rho) =0\ ,
$$
The solution in the inner region is
$$ \phi = i y^{3/2} [J_{3/2} (y) + i N_{3/2} (y) ]
$$
where $y= (\omega R)^3/(2\rho^2)$. This matches onto
$$\phi = 48 \sqrt {2\over \pi} \rho^{-3} J_3 (\rho)
$$
in the outer region. 
The absorption cross section is
found to be \cite{IK,emp}
\beq
\label{two}
\sigma_{\rm 2\ class}= {\pi^4\over 12}\omega^2 R^9= 
{1\over 6\sqrt 2\pi} \kappa_{11}^2 \omega^2 N^{3/2}\ . 
\eeq
 
\subsection{D-brane approach to absorption}

There is a number of minimally coupled scalars in the theory: the
dilaton, the RR scalar, and `off-diagonal'  gravitons polarized with both
indices parallel to the 3-brane world-volume
\cite{GKT}.  In the semiclassical description they obviously have
identical absorption cross-sections. It is remarkable that this is
also true in the D-brane description, and that the D-brane absorption
cross-section is equal to the semiclassical one at low energies
\cite{IK,GKT}. 

The threebrane world-volume theory 
 is ${\cal N}=4$ supersymmetric $U(N)$ gauge theory 
where $N$ is the number of parallel threebranes 
\cite{EW}.  Thus, the massless fields on the
world-volume are the gauge field, 6 scalars, and 4 Majorana fermions,
all in the adjoint representation of $U(N)$. 
The universality of the cross-section is
not trivial in the world-volume description:
while for dilatons and RR scalars leading absorption proceeds by
conversion into a pair of gauge bosons only, for the
gravitons polarized along the brane
it involves a summation over conversions into
world-volume scalars, fermions, and gauge bosons \cite{GKT}.

The world-volume action, excluding all couplings to external fields, is
($I=1,...,4; \ i=4,...,9$)
 \beqs 
   S_3 =& T_{(3)} \int d^4 x \, \tr 
\bigg[ -\tf{1}{4} F_{\alpha\beta}^2 + 
    \tf{i}{2}
 \bar\psi^I \gamma^\alpha \partial_\alpha \psi_I \nonumber \\ &- 
    \tf{1}{2}
 (\partial_\alpha X^i)^2 + {\rm interactions} \bigg] \ .
\eeqs 
We also need to know how the threebrane
world-volume fields couple to the bulk fields of type~IIB
supergravity.  
For the terms necessary to us, the generalization from $U(1)$ to
$U(N)$ gauge group is straightforward. One can, in principle, 
obtain  detailed  information  about the structure of the non-abelian 
action by directly computing
string amplitudes as done in \cite{HK}.  

The leading   bosonic terms in the action for a  single 3-brane 
in a type IIB supergravity background  are \cite{reviews,tse}
\beqs \label{teea}
 S_3&=&-T_{(3)} \int d^4 x \bigg(
    \sqrt {-\det (\hat  g  +  e^{-\p/2} \F) } \nonumber \\
    & +& {\textstyle  {1\ov 4!} } \ep^{\a\b\s\r} \hat C_{\a\b\s\r} 
+   \tf{1}{2}    \hat C_{\a\b} \td \F^{\a\b}
+  \four  C   \F_{\a\b}\td  \F^{\a\b} \bigg) \nonumber
\eeqs
where 
$$ \F_{\a\b}= F _{\a\b}+ \hat B_{\a\b}\ , $$
$$ \td F^{\a\b} = {{\textstyle{1\over 2}}} \ep^{\a\b\s\r}F_{\s\r}\ , 
$$
$$
\hat g_{\a\b} = g_{MN} \del_\a X^M \del_\b X^N \ , {\rm etc.}, 
$$
and the background fields are functions of $X^M$.
In the static gauge ($X^\a = x^\a$, $\a=0,\ldots, 3$) one has 
$$\hat g_{\a\b} = g_{\a\b} + 2g_{i(\a } \del_{\b )} X^i +
 g_{ij } \del_{\a} X^i\del_{\b} X^j 
\ .$$
The additional fermionic terms are dictated by supersymmetry.

The  leading-order 
interaction of the dilaton with world-volume fields 
implied by (\ref{teea}) was first discussed 
in \cite{IK}.  The coupling of the RR scalar $C$  is  
similar, being related by $SL(2,R)$ 
duality. We also include
the coupling of  the gravitons polarized parallel to the brane, 
$h_{\alpha\beta}=g_{\a\b}-\eta_{\a\b}$.

Generalizing to $U(N)$, we find that
the part of $S_{\rm int}$ that is relevant to the leading-order 
absorption processes we wish to consider is\footnote{We  ignore 
the  fermionic couplings like $ \phi  \bar\psi^I \gamma^\alpha
 \partial_{\a} \psi_I $ and  similar ones for $C$ and $h_{\a\b}$ 
 which are proportional to the 
fermionic equations of motion and thus  
give vanishing contribution to the S-matrix elements.}
\beqs \label{sint}
   & S_{\rm int} = T_{(3)} \int d^4 x \, \bigg[ \tr \left(
 \tf{1}{4}   {\phi} F_{\alpha\beta}^2  -
  \tf{1}{4}   {C} F_{\alpha\beta} \td {F}^{\alpha\beta} \right)
\nonumber \\ & +  
     \tf{1}{2} h^{\alpha\beta} T_{\alpha\beta} \bigg] \ , 
 \eeqs 
 where 
\beqs
   & T_{\alpha\beta} = \tr \big[ F_{\alpha}^{\ \gamma} F_{\beta\gamma} - 
    \tf{1}{4} \eta_{\alpha\beta} F_{\gamma\delta}^2 
 - \tf{i}{2} 
\bar\psi^I \gamma_{(\alpha} \partial_{\beta)} \psi_I 
\nonumber \\ &  +
\partial_\alpha X^i \partial_\beta X^i  - 
     \tf{1}{2} \eta_{\alpha\beta} (\partial_\gamma X^i)^2
\big] \ .
 \eeqs 

Let us first consider an off-diagonal graviton
polarized along the brane, say
$h_{xy}$. From (\ref{sint}) one
 can read off the invariant amplitudes
for absorption into two scalars, two fermions, or two gauge bosons
(we have to include $\sqrt 2 \kappa_{10}$, the field normalization
factor for $h_{xy}$).
 Summing over different species and polarizations
of particles available (six
different $X^i$, for example), one obtains
\beqs 
{\rm scalars} &:& \overline{|{\cal M}|}^2 = 
     3 \kappa_{10}^2 \omega^4 n_x^2 n_y^2  \nonumber \\
    {\rm fermions} &:& \overline{|{\cal M}|}^2 = 
     \kappa_{10}^2 \omega^4 (n_x^2 + n_y^2 - 4 n_x^2 n_y^2)  
\nonumber \\
  {\rm vectors} &:& \overline{|{\cal M}|}^2 = 
     \kappa_{10}^2 \omega^4 (1 - n_x^2 - n_y^2 + n_x^2 n_y^2) 
\nonumber  
\eeqs
 where $\vec{n}$ is the direction of one of the outgoing particles.
We have anticipated conservation of energy and momentum by
setting $\vec{p}_1 + \vec{p}_2 = 0$ and $\omega_1 + \omega_2 =
\omega$. 
 It is remarkable that the sum of these three quantities is
independent of $\vec{n}$.  Thus, if one sums
over all the states in the ${\cal N}=4$
super-multiplet, the result is isotropic:
\beq 
   \overline{|{\cal M}|}^2 = \kappa_{10}^2 \omega^4 \ .
\eeq 
 The absorption cross-section is evaluated from 
$\overline{|{\cal M}|}^2$ in precisely the same way 
that decay rates of massive
particles are calculated in conventional 4-dimensional field theories:
 \beqs 
&   \sigma_{3{\rm \ abs}} = {N^2 \over 2} {1 \over 2 \omega} 
    \int {d^3 p_1 \over (2 \pi)^3 2 \omega_1}
         {d^3 p_2 \over (2 \pi)^3 2 \omega_2} \, \nonumber \\
&
    (2 \pi)^4 \delta^4 \big( q - {\textstyle \sum\limits_i} p_i \big) \ 
     \overline{|{\cal M}|}^2 \ .
\eeqs
The leading factor of $N^2$ accounts for the multiple branes and the
$1/2$ is present because the outgoing particles are identical.
The cross-section agrees
with the semi-classical $\ell = 0$ result, (\ref{classical}),
 \beq 
   \sigma_{3{\rm\ abs}} = {\kappa_{10}^2  \omega^3 N^2\over 32 \pi} 
     = \sigma_{3{\rm\ class.}} \ .
 \eeq 
The D-brane calculations for the dilaton and the RR scalar
yield identical results, in agreement with the semiclassical gravity
\cite{IK,GKT}.

It is important to examine the structure of higher power in $g_{\rm
str}$ corrections to the cross-section \cite{Das}.  In semiclassical
supergravity the only quantity present is $\kappa$, and corrections to
(\ref{classical}) can only be of the form
 \beq\label{ACor}
  a_1 \kappa^3 \omega^7+ a_2 \kappa^4 \omega^{11}+ \ldots
 \eeq
 However, in string theory we could in principle find corrections even
to the leading term $\sim \omega^3$, so that
 \beqs\label{BCor}
&\sigma_{3{\rm\ abs}}= {\kappa^2  \omega^3 N^2\over 32 \pi} 
\left (1+ b_1 g_{\rm str} N+ b_2 (g_{\rm str} N)^2+ \ldots \right )
\nonumber \\
&+ {\cal O} (\kappa^3 \omega^7)\ .\eeqs
 Presence of such corrections would spell a manifest disagreement with
supergravity because, as we have explained, the comparison has to be
carried out in the limit $N g_{\rm str}\rightarrow\infty $. Luckily, 
$b_i=0$ due to certain
non-renormalization theorems in $D=4$ ${\cal N}=4$ SYM theory.
In \cite{GKnew} a detailed argument was presented for the absence of such 
corrections in the absorption cross-section of gravitons polarized
along the brane. These gravitons couple to the stress-energy tensor
on the world volume, and it was shown that the absorption cross-section is
determined by the central term in its two-point function. 
Schematically, one finds
  \beq\label{TOPE}
   T(x) T(0) = {c \over x^8} + \ldots \ .
  \eeq
The fact that $b_i=0$ follows from the fact
that the one-loop calculation of the central charge is exact
in $D=4$ ${\cal N}=4$ SYM theory, yielding $c=N^2/4$ 
\cite{AnsOne,AnsTwo,GKnew}.

A more difficult comparison is the absorption cross-section for higher
partial waves. In \cite{IK} the following term was proposed
to describe absorption of a dilaton in the $\ell$-th partial wave, 
\beq 
T_{(3)} \int d^4 x \, \ \tf{1}{4 \cdot \ell!}
    {\partial_{i_1} \cdots \partial_{i_\ell} \phi}
    \tr \left( X^{i_1} \cdots X^{i_\ell}
      F_{\alpha\beta} F^{\alpha\beta}\right)
\eeq 
This term originates from Taylor expansion of the background field
$\phi (X)$ in the world volume action: it describes a process
where an incident dilaton is converted into $\ell+2$ massless
D-brane excitations.
It was shown in \cite{IK} that this term predicts a
cross-section for the $\ell$-th partial wave whose scaling with
$\omega$ and $N$ is in agreement with classical gravity. 
In \cite{GKT} precise agreement in the normalizations was found 
for $\ell = 0,1$, but 
for $\ell>1$ the D-brane cross-section was found to be bigger
than the semiclassical one. 
Some suggestion for how to restore agreement were given in
\cite{GKT}, but the puzzle has not been resolved. This is an
interesting problem for the future.

\subsection{Comparing absorption cross-sections for M-branes}

A remarkable aspect of the agreement between 
the string theoretic and 
the classical
results for threebranes is that it holds exactly 
for any value of  $N$, including
$N=1$. As explained in the introduction, the classical geometry should
be trusted only in the limit $g_{\rm str} N \rightarrow \infty$.
Thus, the agreement of the absorption cross-sections for $N=1$
suggests that our calculations are valid even in the limit
$g_{\rm str} \rightarrow \infty$, provided that $g_{\rm str}
\alpha'^2 \omega^4$ is kept small. The supersymmetric 
non-renormalization theorems are probably at work here, insuring that 
there are no string loop corrections.
We would like to ask whether the exact agreement
between the absorption cross-sections is also found for the twobranes
and fivebranes of M-theory. Their effective actions are not known for
$N>1$, so we can compare the cross-sections only for $N=1$.
In contrast to the threebranes, there is no exact agreement
in the normalizations \cite{GKT}, which
is probably due to the fact that M-theory
has no parameter like $g_{\rm str}$ that can be dialed to make the
classical solution reliable. We will now review some of 
the calculations in \cite{GKT}. 

First we discuss absorption of longitudinally polarized
gravitons by a twobrane. The massless fields in the effective action 
are 8 scalars and 8 Majorana fermions.
The longitudinal
graviton couples to the energy momentum tensor on the
world-volume, $T_{\alpha\beta}$.
The terms in the effective action necessary to describe the absorption of
$h_{xy}$ are ($i= 3, \ldots, 10$;\ $I=1, \ldots, 8$)
\beqs & T_{(2)}
\int d^3 x\  
\bigg [ -\tf{1}{2}  \partial_{\alpha} X^i
\partial^{\alpha} X^i  
+ \tf{i}{2} \bar \psi^I \gamma^\alpha\partial_\alpha
\psi^I\nonumber \\  +&  \sqrt 2 \kappa_{11} 
h_{xy} \big (\partial_x X^i \partial_y X^i 
- \tf{i}{4}  \bar \psi^I
(\gamma_x \partial_y + \gamma_y \partial_x) \psi^I \big ) 
\bigg ]
\ , \nonumber
\eeqs
where $h_{xy}$ is the canonically normalized field which enters the
$D=11$ space-time action as 
$$ - {1\over 2} \int d^{11} x\ \partial_M h_{xy} \partial^M h_{xy} 
\ .$$

The absorption cross-section is found using the Feynman rules in
a way analogous to the threebrane calculation.
For the 8 scalars, we find that the matrix element squared 
(with all the relevant factors included) is
$$ {\kappa_{11}^2 \omega^4 \over 2 } 4 n_x^2 n_y^2
\ ,$$
where $\vec n$ is the unit vector in the direction of one of the
outgoing particles.
For the 8 Majorana fermions, the corresponding object 
summed over the final polarizations is
$$ {\kappa_{11}^2 \omega^4 \over 2 } (n_x^2- n_y^2 )^2 \ . 
$$
Adding them up, we find that the dependence on direction cancels out,
just as in the threebrane case. The sum must be multiplied by
the phase space factor ${1\over 8 \omega^2}$,
so that the total cross-section is
\beq
\sigma_{2{\rm \ abs}} = {\kappa_{11}^2 \omega^2 \over 16 } 
\ .
\eeq
This does not agree with the classical result 
(\ref{two}) for $N$ set to 1.
Notice that even the power of $\pi$ does not match.
This situation is reminiscent of the
discrepancy in the near-extremal entropy where the relative factor was 
a transcendental number involving $\zeta(3)$ \cite{kt}. 

Now we turn to the fivebrane.
The massless fields on the fivebrane form a tensor
multiplet consisting of 5 scalars, 2 Weyl fermions and the
  antisymmetric tensor $\B_{\alpha\beta}$  with anti-selfdual 
strength.
To discuss the absorption of longitudinally polarized 
gravitons, $h_{xy}$, we need the action ($i=6, \ldots, 10$; \ 
$I=1, 2$)
$$ T_{(5)}
\int d^6 x\  
\bigg [ - \ \tf{1}{2} \partial_{\alpha} X^i
\partial^{\alpha} X^i - \tf{1}{12} \H_{\alpha\beta\gamma}^2 
+i \bar \psi^I \gamma^\alpha\partial_\alpha
\psi^I
$$
\beqs  
&+& \sqrt 2 \kappa_{11} h_{xy} 
\big (\partial_x X^i \partial_y X^i 
+  \tf{1}{2} \H^-_{x\beta\gamma} \H_y^{- \beta\gamma}
\nonumber \\ &-&  \tf{i}{2} \bar \psi^I
(\gamma_x \partial_y + \gamma_y \partial_x) \psi^I \big)
\bigg ]
\ . 
\eeqs
For the 5 scalars, we find that the matrix element squared 
(with all the relevant factors included) is
\beq
{\kappa_{11}^2 \omega^4 \over 2 } {5\over 2} n_x^2 n_y^2
\ .
\eeq
For the 2 Weyl fermions, the corresponding object 
summed over the final polarizations is
\beq
{\kappa_{11}^2 \omega^4 \over 2 } (n_x^2+ n_y^2 - 4 n_x^2 n_y^2 )
\ .
\eeq
Finally,
the contribution  of the anti-selfdual gauge field
turns out to be equal to that  of the
usual, unconstrained 
$\H_{\a\b\g}$  divided by 2.
Summing over polarizations, we find, 
\beq
{\kappa_{11}^2 \omega^4 \over 2 } 
\left (1- n_x^2- n_y^2 + {3\over 2} n_x^2 n_y^2\right ) 
\ .
\eeq

Adding up the contributions of the entire tensor multiplet, we find
that all the direction-dependent terms cancel out, just as they did
for the threebrane and the twobrane.
Multiplying by the phase space factor 
${\omega\over 2^7\cdot 3\pi^2}$, we find that 
the total cross-section for the longitudinally
polarized gravitons is 
\beq
\sigma_{5{\rm \ abs}}={\kappa_{11}^2 \omega^5\over 2^8 \cdot 3 \pi^2}
\ .\eeq
This turns out to be
a factor of 4 smaller than the classical result (\ref{five})
evaluated for $N=1$.  

What is the reason for this discrepancy?
For $N=1$, the curvature of the 
solution is of order of the 11-dimensional Planck scale.
Obviously, the 11-dimensional supergravity is at best a low-energy
approximation to M-theory. The M-theory effective action should contain
higher-derivative terms weighted by powers of $\kappa_{11}$, by
analogy with the $\alpha'$ and $g_{\rm str}$  
expansions of the string effective action.
Thus, for $N=1$, the classical solution may undergo corrections of order
one which we believe to be the source of the discrepancy.
For large $N$, however, we expect the M-theory cross-section to agree
exactly with the classical cross-section.
On the other hand, the absorption cross-section for gravitons 
polarized along the brane is in general related to the Schwinger term in
the two-point function of the stress-energy tensors \cite{GKnew}.
The form of the semiclassical cross-section for $N$ M2-branes
(\ref{two}) indicates that we are dealing with a superconformal field theory
in 2+1 dimensions. Schematically, the OPE is
\beq
   T(x) T(0) = {c_2 \over x^6} + \ldots \ ,
  \eeq
and the behavior of the
central charge is $c_2\sim N^{3/2}$ for large $N$ \cite{GKnew}.
For $N$ M5-branes the semiclassical cross-section 
(\ref{five}) indicates that we are dealing with a superconformal field theory
in 5+1 dimensions. Now the schematic OPE of stress-energy tensors is
\beq
   T(x) T(0) = {c_5 \over x^{12}} + \ldots \ ,
  \eeq
and the behavior of the
central charge is $c_5\sim N^3$ for large $N$ \cite{GKnew}.

These results have an obvious connection with properties of
the near-extremal entropy found in \cite{kt}.
Indeed, the near-extremal entropy of a large number $N$
of coincident M2-branes is formally
reproduced by ${\cal O}(N^{3/2})$
massless free fields in 2+1 dimensions, while that of
$N$ coincident M5-branes is reproduced by ${\cal O}(N^3)$
massless free fields in 5+1 dimensions.

\section{Creation of Fundamental Strings by Crossing D-branes}

The D-branes are BPS saturated objects which preserve
16 supersymmetries out of the original 32. 
This implies that when two Dirichlet $p$-branes are
placed parallel to each other, the force between them vanishes.
A string theoretic calculation of this force involves the cylinder
diagram, with the ends of the cylinder attached to different D-branes.
In the open string channel there are contributions from the 
NS, R, and NS $(-1)^F$ sectors, which cancel due to the abstruse
identity for theta-functions \cite{polch}. Physically, this means that the
attraction due to NS-NS closed strings, the graviton and the dilaton,
is canceled by the repulsion of the like R-R charges.

It is interesting to study a more general situation where a
$p'$-brane is placed parallel to a $p$-brane with $p'<p$.
This configuration preserves 8 of the supersymmetries if 
$p-p'=4$ or 8 \cite{reviews}. 
For $p-p'=4$ only the NS and the R open string sectors
contribute to the cylinder amplitude, and they cancel identically.
Thus, there is no force due to the R-R exchange, while the graviton and
the dilaton forces cancel identically.

A more complicated situation
arises for $p-p'=8$. One example of this is the 0-brane
near the 8-brane, which is important for understanding the heterotic
theory \cite{DF}. 
Now the contribution of the NS and R open string sectors to
the cylinder amplitude is \cite{Lif}
\beqs
& A_{{\rm NS-NS}}= {1\over 2}\int_0^\infty {dt\over t}
(8\pi^2\alpha' t)^{-1/2} e^{-{t Y^2\over 2\pi\alpha'}}
f_4^{-8}(q)\times \nonumber \\ 
& \left (-f_2^8 (q)+ f_3^8(q) \right ) =
{1\over 2}\int_0^\infty {dt\over t}
(8\pi^2\alpha' t)^{-1/2} e^{-{t Y^2\over 2\pi\alpha'}}
\nonumber \eeqs
where $q=e^{-\pi t}$ and $Y$ is the transverse position of the
0-brane relative to the 8-brane.
Thus, we find a constant repulsive force due to the NS-NS closed
strings,
\begin{equation} \label{repulsion}
-{\partial A_{{\rm NS-NS}}(Y)\over \partial Y}=
{1\over 4\pi\alpha'} {\rm sign}(Y)\ .
\end{equation} 
As pointed out by Lifschytz \cite{Lif},
this is canceled by a contribution of the R $(-1)^F$ open string
sector, which implies that there is attraction due to R-R exchange.
The nature of this attraction was elucidated in \cite{DFK}.

A peculiar feature of this force is that
it jumps by $\pm {1\over 2\pi \alpha'}$ every time the
0-brane crosses the 8-brane.
This jump is due to {\bf creation} of a
fundamental string stretched between the 0-brane 
and the 8-brane \cite{DFK}.
This phenomenon is similar, and in fact U-dual, to the creation of
a 3-brane discovered by Hanany and Witten \cite{HW}. Since the number of
stretched fundamental strings jumps by $\pm 1$ upon each crossing,
we may regard the ground state of the 0-8 system as containing
$\pm {1\over 2}$ of a fundamental string (the sign refers to whether
the string enters or exits the 0-brane).
When the 0-brane is to the left of the
8-brane, we have, say, $-{1\over 2}$ of a fundamental string.
Upon crossing, this turns into $+{1\over 2}$. 
The attractive force equal to ${1\over 2}$ of the
fundamental string tension is what is necessary to cancel the repulsion
due to the graviton and the dilaton,
(\ref{repulsion}). This is how the no-force
condition required by supersymmetry is maintained in the 0-8 system.

\subsection{U-duality and creation of a fundamental string}

The creation of a stretched string by
a 0-brane crossing an 8-brane is related by U-duality to creation
of a stretched 3-brane by a R-R 5-brane crossing a NS-NS 5-brane.
Hanany and Witten showed that, when an R-R charged 5-brane 
positioned in the $(1-2-6-7-8)$ directions
crosses a NS-NS charged 5-brane positioned in the $(1-2-3-4-5)$ 
directions, a single $(1-2-9)$ 3-brane stretched between the 5-branes
is created \cite{HW}.

Applying T-duality along directions 1 and 2 we find that,
when a $(6-7-8)$ 3-brane crosses a $(1-2-3-4-5)$ NS-NS 5-brane,
then a D-string stretched between them along the 9th direction is
created. From the S-duality of the type IIB theory it now follows
that, when a $(6-7-8)$ 3-brane crosses a $(1-2-3-4-5)$ R-R 5-brane,
then a fundamental string stretched between them along the 9th direction
is
created. This is the kind of process that is of primary interest to us,
because it involves two D-branes with 8 ND coordinates. There are a number
of other such processes related to this by T-duality. For example,
after T-dualizing along directions 6, 7 and 8, we find that a
0-brane crossing an 8-brane creates a stretched fundamental string.

It is interesting that both the Hanany-Witten process and the
fundamental string creation originate from the same phenomenon in
M-theory: creation of a 2-brane by crossing 5-branes.
Indeed, when a $(2-3-4-5-10)$ 5-brane crosses a
$(6-7-8-9-10)$ 5-brane, a $(1-10)$ 2-brane stretched between the 5-branes
is
created. Reducing to the type IIA theory along direction 5, we find that
a 4-brane crossing a 5-brane creates a 2-brane. This is T-dual
to the 3-brane creation discussed in \cite{HW}. 
We may, however, choose to reduce to the type IIA theory along direction
$10$, which is common to all the branes. Then we find that 
a $(2-3-4-5)$ 4-brane crossing a $(6-7-8-9)$ 4-brane
creates a fundamental string stretched along direction 1.
This confirms that two crossing D-branes, positioned in such a way
that there are 8 ND coordinates, create a stretched fundamental string.

\subsection{Effective action arguments}

Let us give a direct argument for the creation of
fundamental strings. 
For concreteness, we will refer to the 0-8 system, but
analogous arguments apply to all cases related to this by T-duality.

The term in the 8-brane world volume action which is crucial for our 
purposes is \cite{polch}
\begin{equation} \label{csterm}
\mu_{(8)} {1\over 2\cdot 7!} \int d^9 \sigma
\epsilon_{\nu_0 \ldots \nu_8} C_{(7)}^{\nu_0 \ldots \nu_6}
F^{\nu_7 \nu_8}\ ,
\end{equation} 
where $C_{(7)}$ is an R-R potential, and
$F= dA$ is the world volume gauge field strength.
The D-brane charge densities were determined in \cite{polch} to be
$$ \mu_{(p)}= \sqrt{2\pi} (2\pi \sqrt{\alpha'})^{3-p}
\ .
$$
Integrating (\ref{csterm}) by parts, we get
\beqs \label{newcsterm}
& {\mu_{(8)} \over 8!} \int d^9 \sigma
\epsilon_{\nu_0 \ldots \nu_8} F_{(8)}^{\nu_0 \ldots \nu_7}
A^{\nu_8} \nonumber \\
&= \mu_{(8)} \int d^9 \sigma F_{(2)}^{\mu 9} A_\mu\ ,
\eeqs 
where
$$ F_{(8)}= d C_{(7)}\ , \qquad F_{(2)}= ^* F_{(8)}
\ ,$$
and $9$ is the direction normal to the 8-brane.

In the presence of a stationary 0-brane, there is a radial 
electric field,
\beq
F_{(2)}^{0r}= {\mu_{(0)}\over r^8 \Omega_8}\ ,
\eeq
where $\Omega_8$ is the volume of a unit 8-sphere.
Eq. (\ref{newcsterm}) shows that the normal component of the electric
field, $F_{(2)}^{09}$, plays the role of the charge
density in the world volume gauge theory. The total charge on the
8-brane is
\beq
\mu_{(8)} \int d^8 \sigma F_{(2)}^{09}= {1\over 2}
\mu_{(8)} \mu_{(0)}= {1\over 4\pi \alpha'}\ .
\eeq
Let us recall that an endpoint of a fundamental string manifests
itself in the world volume gauge theory
as an electric charge of magnitude $\pm {1\over 2\pi \alpha'}$.
We conclude that the 0-brane and the 8-brane are connected
by {\bf one half} of a fundamental string.
This provides the attraction that cancels the repulsion from the
graviton-dilaton exchange.

As the 0-brane crosses the 8-brane, the net electric charge on the 
8-brane jumps from ${1\over 4\pi \alpha'}$ to
$- {1\over 4\pi \alpha'}$. This clearly shows that an endpoint of
a fundamental string is created on the 8-brane. Similar considerations
in the 0-brane action show that
the other end of the string is attached to the 0-brane. 
The term in the 0-brane action responsible for this effect is 
\begin{equation} \label{strangeterm}
\mu_{(0)}\int d\tau F A_0\ ,
\end{equation} 
where $F=^* F_{(10)}$ is the zero-form field strength dual
to the 10-form emitted by the 8-brane. Thus, 
$\mu_{(0)} F$ is the `source' for $A_0$. We believe that this shows that
the fundamental string indeed ends on the 0-brane.\footnote{
A similar conclusion was reached in \cite{PS,bgl}. 
In the presence of an 8-brane the type IIA supergravity is
massive, and the equations of motion imply
that a string must end on a
0-brane. A careful normalization shows that, when the 0-brane
is to the right (left) of the 8-brane, $-{1\over 2} \left
(+{1\over 2}\right )$ of a fundamental string ends on
the 0-brane. Brane creation has also been analyzed using the anomaly 
inflow argument \cite{badogr}.}
Correctness of this argument may be checked through T-duality.
For instance, if we T-dualize the 0-8 system to a pair of orthogonal
4-branes,
then (\ref{strangeterm}) goes into the following term of the 4-brane
action,
\begin{equation} 
{\mu_{(4)}\over 4!}\int d^5\sigma 
\epsilon_{\nu_0 \ldots \nu_4} F_{(4)}^{\nu_0 \nu_1 \nu_2 \nu_3 }
A^{\nu_4}\ .
\end{equation}
The jump in the total charge on a 4-brane as it is crossed by
the other 4-brane is 
$$ \mu_{(4)}^2 ={1\over 2\pi\alpha'}\ ,
$$ 
which is precisely the tension of one fundamental string. 

The `bare' Chern-Simons term (\ref{strangeterm})
is crucial for the consistency
of the quantum mechanics of a 0-brane near a 8-brane.
The 0-8 strings are described by a complex fermion $\chi$ with
the lagrangian
$$ L_\chi= -i \bar\chi \dot \chi -
\bar\chi Y \chi - \bar\chi A_0 \chi
\ ,$$
where the scalar $Y$ is the distance between the 0-brane and the
8-brane (we are now working in the units where $2\pi \alpha'=1$).

If we integrate out the fermions, we find 
`induced' Chern-Simons and potential terms \cite{BSS}
\begin{equation} \label{induced}
 {1\over 2} {\rm sign}(Y) (Y+ A_0)
\ .
\end{equation}
Thus, the theory seems anomalous because the coefficient in front
of the Chern-Simons term is fractional. This problem is solved by
the creation of the fundamental strings. The `bare' Chern-Simons term
(\ref{strangeterm}),
$$ \mu_{(0)} F A_0 = -{1\over 2} {\rm sign}(Y) A_0
\ ,$$
exactly cancels the `induced' term in (\ref{induced}).
The `bare' term
may be interpreted as due to $1/2$ of a fundamental
string. There is a corresponding
`bare' attractive potential related to (\ref{strangeterm})
by $N=8$ supersymmetry \cite{BSS}, 
$$\delta L=-{1\over 2} |Y| \ ,$$
which cancels the `induced' repulsive potential.
Thus, the entire `induced' lagrangian (\ref{induced}) obtained by
integrating over the fermions is canceled by the `bare' terms
due to $1/2$ of a fundamental string.

In the previous section we showed that the string creation follows
by dimensional reduction from membrane creation in M-theory.
Let us make a direct argument for the latter. Consider the effective
action for a $(1-2-3-4-5)$ 5-brane in the presence of a
$(1-6-7-8-9)$ 5-brane. This action contains a Chern-Simons term
\begin{equation} \label{fivecsterm}
q_{(5)} {1\over (3!)^2} \int d^6 \sigma
\epsilon_{\nu_0 \ldots \nu_5} C^{\nu_0 \nu_1 \nu_2}
H^{\nu_3 \nu_4 \nu_5}\ ,
\end{equation} 
where $H= dB$ is the world volume field strength.
The 2-brane and 5-brane charge densities and tensions were normalized
in \cite{KT},
\begin{equation}
q_{(2)} =\sqrt 2 \kappa T_{(2)} =   \sqrt 2 (2\kappa \pi^2)^{1/3} \ , 
\end{equation}
\begin{equation}
 q_{(5)} =\sqrt 2 \kappa T_{(5)} =  \sqrt 2 ({\pi\over 2\kappa})^{1/3} \ .
\end{equation}
Integrating (\ref{fivecsterm}) by parts, we find
\begin{equation} \label{newfivecsterm}
q_{(5)} {1\over 2\cdot 4!} \int d^6 \sigma
\epsilon_{\nu_0 \ldots \nu_5} F^{\nu_0 \nu_1 \nu_2 \nu_3 }
B^{\nu_4 \nu_5}\ .
\end{equation} 
This shows that $F_{2345}$ acts as a source for $B_{01}$.
Thus, $F_{2345}$ is proportional to the density of strings on the world
volume which point along direction 1. Such a string is the boundary
of a $(1-10)$ 2-brane stretched between the 5-branes.
Evaluating the flux through the $(1-2-3-4-5)$ 5-brane due to the 
$(1-6-7-8-9)$ 5-brane,
$$ \int d^5 \sigma F_{2345}\ ,
$$
we find that the net charge that couples to $B_{01}$ is
\begin{equation} \label{chargerel}
 { q_{(5)}^2\over 2} = {T_{(2)}\over 2}
\ .
\end{equation}
Thus, {\bf one half} of a 2-brane is stretched between the
5-branes. As the 5-branes pass through each other, one 2-brane
is created. It is interesting that this process is encoded in
the relation (\ref{chargerel}) between the charge
of the 5-brane and the tension of the 2-brane in M-theory.

\subsection{Implications for the type I' theory}

The phenomena we have discussed have interesting implications
for the physics of the type I' theory, which is S-dual to the heterotic
string. Its set-up involves 16 D8-branes located between two
8-orientifold planes. The dynamics of the zero-branes is
described by the quantum mechanical system introduced in \cite{DF}. 
At first sight it appears that there
exists an effective linear potential for the 0-brane which
experiences a discontinuity in the slope every time a 0-brane
passes through an 8-brane. This potential, however, results from
including only the NS-NS exchange in the string theory cylinder
diagram. As we have shown, inclusion of the R-R term interpretable as
the string creation cancels the discontinuity in the force. In fact,
the entire linear potential cancels everywhere \cite{DFK}.
Similarly the net
Chern-Simons term also cancels, as required by the $N=8$ supersymmetry.
This is consistent with the picture that we 
have developed in the preceding sections.

\section*{Acknowledgments}

These notes are based on conference talks delivered at 
Strings '97 in Amsterdam and at SUSY '97 in Philadelphia.
I thank the organizers of these conferences for their
hospitality. I am grateful to U. Danielsson, G. Ferretti,
S. Gubser and A. Tseytlin for collaboration on the material
presented here.
This work was supported in part by the DOE grant DE-FG02-91ER40671,
the NSF Presidential Young Investigator Award PHY-9157482, and the
James S.{} McDonnell Foundation grant No.{} 91-48.



\end{document}